\documentclass{appolb}
\usepackage{graphicx}

\begin{document}
\title{Manifestation of pairing modes in nuclear collisions %
\thanks{Presented at Zakopane Conference on Nuclear Physics: "Extremes of the Nuclear Landscape", 2022}%
}
\author{A. Makowski$^{a}$, M.C. Barton$^{a}$, P. Magierski$^{a,b}$, K. Sekizawa$^{c,d}$ and G. Wlaz{\l}owski$^{a,b}$
\address{$^a$ Faculty of Physics, Warsaw University of Technology, Poland\\ 
$^b$ Department of Physics, University of Washington, USA\\
$^c$ Department of Physics, School of Science, Tokyo Institute of Technology, Japan\\
$^d$ Nuclear Physics Division, Center for Computational Sciences, University of Tsukuba, Japan}}

\maketitle
\begin{abstract}
We discuss the possible manifestation of pairing dynamics in nuclear collisions beyond the standard quasi-static treatment of pairing correlations. 
These involve solitonic excitations induced by pairing phase difference of colliding nuclei and pairing dynamic enhancement in the di-nuclear system formed by
merging nuclei.
\end{abstract}

\vspace{-20pt}
\section{Introduction}
Pairing correlations play a crucial role in our understanding of the properties of nuclear systems, ranging from atomic nuclei to neutron stars~\cite{BCS:2013}. 
The importance of pairing correlations, however, do not originate from their contribution to the energy of nuclear systems.
Indeed the pairing energy is only a small fraction of the total energy of an atomic nucleus. 
This is because the value of a pairing gap, which sets the typical energy scale, does not exceed $3\%$ of Fermi energy. 
At subnuclear densities, characteristic for the neutron star crust, it may reach at most about $5\%$.
The importance of pairing correlations lies in the modification induced at the Fermi surface, which produces a gap in the single particle spectrum.
Consequently, it facilitates large amplitude nuclear motion by suppressing dissipative effects due to single-particle excitations.
Thus, the main effect originates from the gap size, which is a single number associated with Cooper pair correlation energy and can be generated within BCS theory~\cite{ringschuck}. 
This description is satisfactory if one describes a situation close to an adiabatic limit of nuclear motion.
In this case, one effectively describes quantum evolution as going through almost static solutions obtained within the static BCS equations.
In the extreme limit of cranking approximation, the evolution of pairing is just provided by instantaneous static gap values, and the pairing gap is simply a function of collective variables describing large amplitude nuclear motion.

The question which naturally arises is whether this approach is always correct.
Recent theoretical investigations of pairing dynamics indicated that even when a nucleus evolves slowly during the fission process, its motion hardly fulfils adiabatic criteria and the pairing field fluctuates rapidly in time and space~\cite{bulgac16}.
Therefore, it is crucial to specify the conditions where the adiabatic approach must be abandoned and to understand possible manifestations of pairing dynamics (see eg. Refs. \cite{bulgac16, magierski2017, magierski2022, scamps_dynamics, tong_dctdhfb} 
for the description of pairing beyond the adiabatic approximation).

\vspace{-20pt}
\section{Nuclear collisions and pairing dynamics}
Nuclear processes expected to elude adiabatic description are nuclear collisions, even at energies close to the Coulomb barrier.
The best examples are provided by nuclear collisions of medium mass nuclei or those involving heavy targets. 
The latter ones are essential in superheavy element synthesis~\cite{hofmann}.

\begin{figure}[t]
\centering
		\includegraphics[width=0.85\columnwidth]{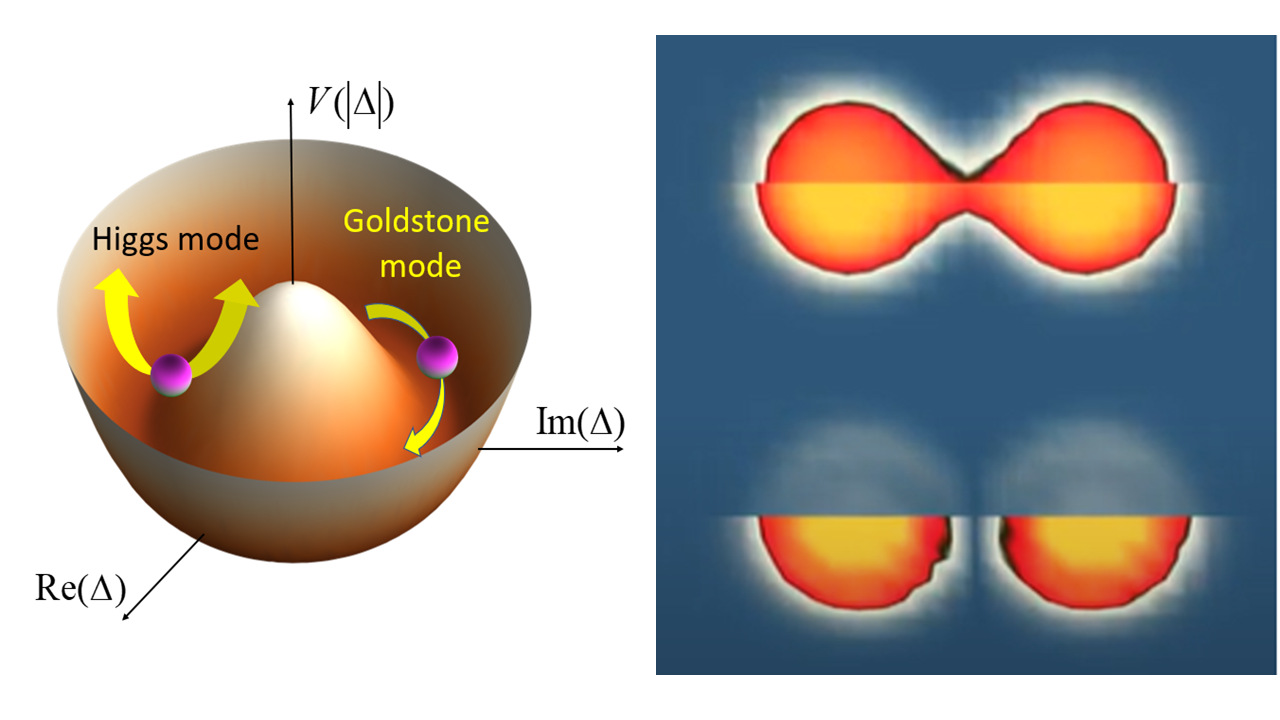}
		\caption[hg]{{\bf Left}: Schematic figure showing Goldstone and Higgs modes associated with symmetry breaking due to emergence of pairing. 
		  {\bf Right}: snapshot from TDDFT simulation of $^{96}$Zr + $^{96}$Zr collision at $E_{cm}=187$ MeV with opposite phases of pairing fields. 
		  The upper panel shows density distributions for protons (upper part) and neutrons (lower part). 
		  The lower panel shows an analogous magnitude of pairing field distributions with solitonic excitation visible between colliding nuclei.
		  Details of calculations are presented in Ref.~\cite{magierski2022}.}
		\label{fig:hg1}
	\end{figure}

What can one expect concerning pairing dynamics in the case of a collision?
We describe pairing as a pairing field constituting an order parameter emerging from U(1) symmetry breaking. 
In that case two obvious possibilities arise, defined by two fundamental modes associated with pairing: Goldstone mode and Higgs mode (see Fig.~\ref{fig:hg1}).
They are associated with variations in the phase and the magnitude of the pairing field, respectively.
Goldstone mode, in its most direct realization, leads to harmonic vibration of phase $\phi({\bf r}, t) \propto {\bf k}\cdot{\bf r} - \omega t$, giving rise to Anderson-Bogoliubov phonons~(see eg. \cite{hoinka,inakura} and references therein).
However, in the atomic nucleus, due to its small size, such modes cannot be unambiguously defined\footnote{Situation is slightly different in neutron star crust where such modes are predicted to show up in neutron superfluid surrounding nuclear impurities~\cite{inakura}.}.
The manifestation of the Goldstone mode can also appear due to perturbation of the nuclear pairing phase induced by dynamics of collision. 
This situation may occur in two regimes. 
The first one corresponds to the case when two nuclei approach each other at subbarrier energies, and the effective phase difference of their pairing fields induces tunnelling of nucleons~\cite{josephson}.
When a collision occurs above the barrier, solitonic excitation is generated between colliding nuclei (see Fig. 1)~\cite{magierski2017}.
These two regimes have been identified and studied in ultracold atomic gases~\cite{valtolina}.
In nuclear systems, the first one has been investigated as a nuclear manifestation of the Josephson effect.
Recently it has been found that an oscillating flow of neutrons occurs (analogue of AC Josephson junction) during a collision of medium-mass nuclei~\cite{potel21}. 
The other regime, leading to solitonic excitation, has been identified in Ref.\cite{magierski2017}.
The difference between these two regimes lies in the expected outcomes. In the first case the main observable is the enhanced nucleon transfer, whereas in the other regime, one expects an additional energy barrier preventing the merging of colliding nuclei.
This additional energy barrier scales with the phase difference between colliding nuclei  $\Delta\phi$ like $\sin^{2} (\Delta\phi/2)$, which was confirmed in TDDFT calculations~\cite{magierski2017}.

\begin{figure}[t]
\centering
		\includegraphics[width=0.95\columnwidth]{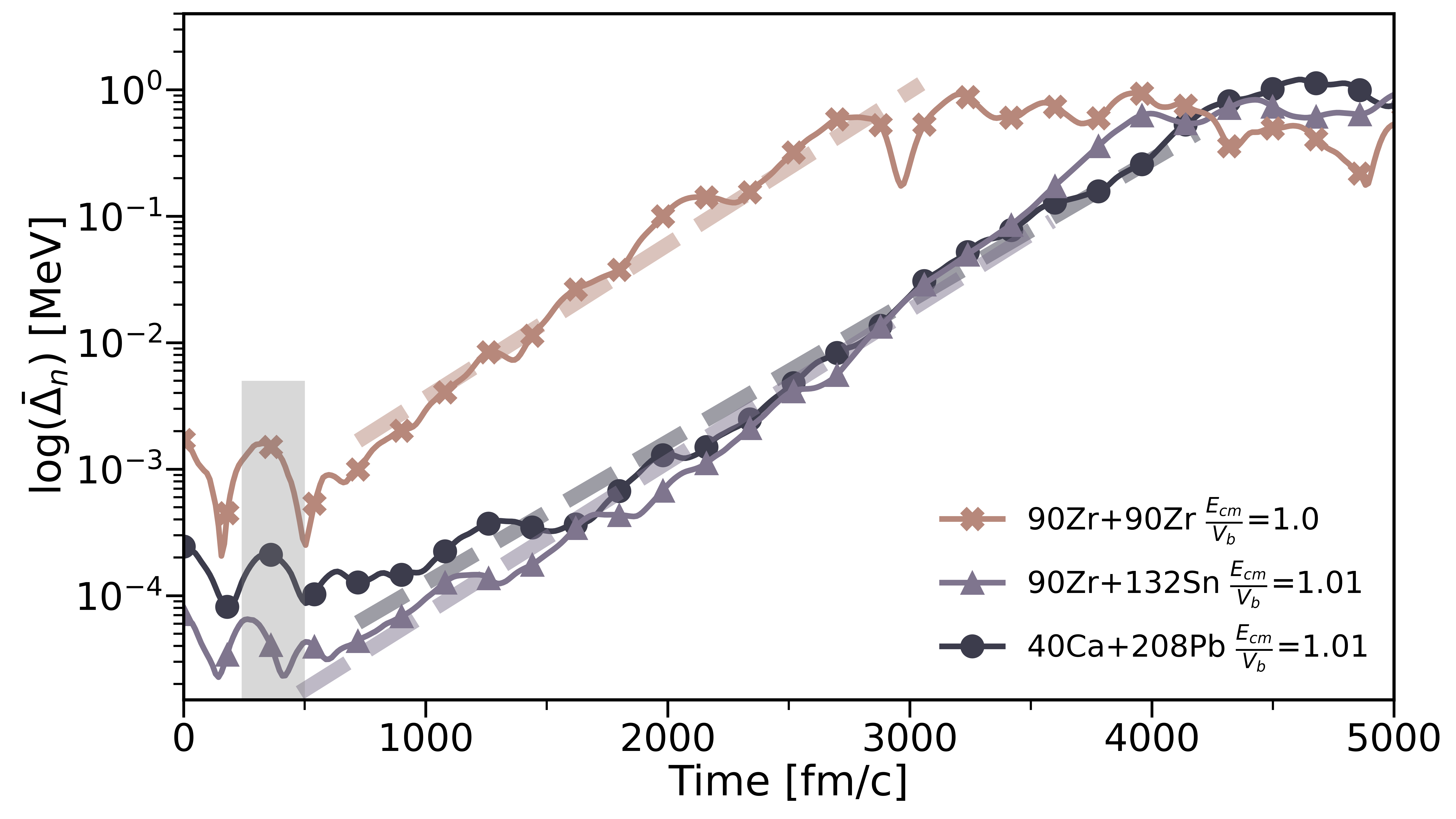}
		\caption[hg]{Magnitude of an average neutron pairing gap $\bar{\Delta}_{n}$ in collisions of several neutron magic nuclei at energies right above the Coulomb barrier ($E_{cm}$ to static barrier ratio is shown in the legend).
		The collision occurs at about $t\approx400$\,fm/$c$. 
		Details of calculations are presented in Ref.~\cite{magierski2022}.}
		\label{fig:hg2}
\end{figure}

Spontaneous symmetry breaking generates also another effect, which leads to pairing magnitude vibrations.
It can be generated in ultracold Fermi gas by tuning the coupling constant in real time, which drives the system towards the superfluid phase~\cite{behrle}.
The characteristic feature of the Higgs mode is its energy (or frequency of oscillations), which is of the order of the static value of the pairing gap.
At first it may seem that this mechanism cannot operate in nuclear systems as pairing correlations emerge from nuclear interaction and cannot be tuned at will. 
However, the effective strength of interaction depends on the density of 
states at the Fermi surface.
This can undoubtedly change once the nuclear shape evolves. 
In particular, when two nuclei merge, a new system is formed during a nuclear collision. 
The single particle properties of such, usually very elongated object, are significantly different from those of two initial nuclei.
It is, therefore, possible that effectively merging two nuclei creates a system which exhibits pairing instability~\cite{magierski2022,magierski2023}.
This would correspond to an exponential increase of the strength of pairing correlations in time. 
This is indeed the case, as seen in Fig. \ref{fig:hg2}.
In the figure the quantity $\bar{\Delta}_{n}=\frac{1}{N}\int d^{3}r |\Delta_{n}({\bf r})|\rho_n({\bf r})$ has been shown as a function of time. 
Here, $\rho_n$ and $\Delta_{n}$ stand for the neutrons density and pairing field distributions, respectively. 
$N=\int \rho_n({\bf r})\,d^{3}r$ is total number of neutrons. 
The initial pairing of colliding nuclei is very weak but become strongly enhanced after collision showing clearly an instability as indicated by almost perfectly exponential growth.
Although it is tempting to associate this effect with the excitation of a Higgs mode, one has to be careful. 
First, the time scale of the enhancement is by an order of magnitude longer than the typical time scale of Higgs mode which has to be comparable to $\hbar/\bar{\Delta}_{n} \approx 200$\,fm/$c$. 
Second, the excitation energy of the system is rather high. 
Using Thomas-Fermi approach~\cite{magierski2022} one may estimate the excitation energy related to neck formation between two nuclei during collision.
For the reaction presented in Fig.~\ref{fig:hg2} it reads: $20,\ 27,\ 34$ MeV for $^{90}$Zr+$^{90}$Zr,\ $^{90}$Zr+$^{132}$Sn and $^{40}$Ca+$^{208}$Pb, respectively.
These energies correspond to temperatures which are close to critical temperature.  
Therefore it seems unlikely to associate such a mode with inducing an actual superfluid phase and it is rather related to the increase of pairing correlations in a nonequilibrium system~\cite{magierski2023}.

\vspace{-15pt}
\section{Conclusion}
We have discussed two examples of the manifestation of pairing dynamics which are predicted to occur in nuclear collisions at the energies close to the Coulomb barrier. 
It is essential to make a systematic assessment of the importance of these effects and to understand their role in nuclear dynamics, particularly in the quasifission process.

\vspace{-15pt}
\section*{Acknowledgments}
We want to thank Nicolas Schunck and his collaborators for help concerning the usage of HFBTHO (v4.0)~\cite{nicolas_hfbtho}. This work was supported by the Polish National Science Center (NCN) under Contracts No. UMO-2017/27/B/ST2/02792.
We acknowledge the support of Global Scientific Information and Computing Center, Tokyo Institute of Technology for resources at TSUBAME3.0 (project ID: hp220072).

\end{document}